# Large-scale transfer and characterization of macroscopic periodically nano-rippled graphene

*Iva Šrut Rakić[1*], Davor Čapeta[2], Milivoj Plodinec[3], Marko Kralj[1].*

[1]Institut za fiziku, Bijenička 46, 10000 Zagreb, Croatia,

[2]Faculty of science, University of Zagreb, Bijenička 32, 10000 Zagreb, Croatia

[3]Institut Ruđer Bošković, Bijenička cesta 54, 10000 Zagreb, Croatia

**Abstract**

Nano-rippled graphene, a structurally modified graphene, presents a novel material with a large range of possible applications including sensors, electrodes, coatings, optoelectronics, spintronics and straintronics. In this work we have synthesized macroscopic single layer graphene with well-defined uniaxial periodic modulation on a stepped Ir(332) substrate and transferred it to a dielectric support. The applied fast transfer process does not damage the Ir crystal which can be repeatedly used for graphene synthesis. Upon transfer, a millimeter sized graphene flake with a uniform periodic nano-ripple structure is obtained, which exhibits a macroscopically measurable uniaxial strain. The periodic one dimensional arrangement of graphene ripples was confirmed by atomic force microscopy and polarized Raman measurements. An important feature of this system is that the graphene lattice is rotated in several different, well-defined orientations with respect to the direction of the ripple induced uniaxial strain. Moreover, geometry of the ripples can be modified by changing the graphene synthesis parameters.

## 1. Introduction

Ever since its discovery in 2004 [1] graphene has presented itself useful for diverse applications [2–5], and has sprang many variants with different properties such as bilayer graphene, nanoribbons, nanomesh and nano-rippled graphene (NRG). Nano-rippled graphene is a representative of structurally modified graphene and presents a novel material adequate for a large range of possible applications including three dimensional (3D) electrodes for batteries, advanced coatings, sensors, optoelectronics, spintronics and straintronics. In a broader context, structurally modified graphene belongs to a class of carbon-based nanomaterials of improved properties, which are successfully designed for a number of different applications, e.g. coatings, composite fibers or quantum dots applications [6–8]. The possibility of structural modification of graphene and periodic texturing to form rippled structures, relies on the ability of this exceptional material to withstand large deformations [9,10]. Physical modulation of graphene increases its surface area and affects properties such as chemical reactivity [11]. Moreover, large deformations and strain in graphene are related to the effects such as large effective magnetic fields [12], electronic band structure changes, a gap opening [13–17] as well as changes in the conductance [18,19] and Raman active modes [20]. All of this leads to an array of theoretically predicted and experimentally realized applications in strain mediated electronics (straintronics) [21,22], optoelectronics [23,24], various sensors [25,26], field emitters [27], and even spintronics [28].

[*]Corresponding author.   Tel: +3851 469 8840.   E-mail: isrut@ifs.hr

Nano-rippled graphene can be intrinsically occurring, for instance in a suspended form [29,30], but also in structures such as wrinkles formed during the epitaxial growth [31], ripples that emerge during transfer process, *e.g.*, due to existence of intrinsic nonperiodic substrate roughness [32], steps on the supporting substrate [27,28,33], or in graphene transferred to a prestrained surface [25]. More recently, large anisotropic graphene crumpling was demonstrated over millimeter sized graphene areas via thermally induced contractile deformations [34], However, in most of these cases, ripples are sporadic or quasi-periodic at best, nonuniform over millimeter scale, with poor control over their geometry and orientation, which is not favorable for applications in straintronics [14,15]. A more restrained approach was suggested by transferring graphene to a prestructured surface [21] where a better control over periodicity was achieved, however, the issue of perfect alignment remains open, and for some applications corrugated supports are not desirable.

In this paper we present a new way to obtain NRG on a dielectric support which is periodically modulated in one dimension (1D), with the modulation equally aligned over a macroscopic scale. We have synthesized millimeter-sized single layer graphene on a stepped Ir(332) surface (gr/Ir(332)) and subsequently transferred it to a SiO$_2$ terminated Si (Si/SiO$_2$) substrate implementing a modified bubbling transfer method [35,36]. This fast transfer process leaves the Ir crystal undamaged and the size of the transferred graphene flake is only limited by the size of the Ir crystal itself. The prominent feature of the transferred graphene is that it keeps the characteristic rippling of the original epitaxial sample and displays a parallel array of ripples over the entire sample. Moreover, such graphene exhibits a macroscopically detectable inherent uniaxial strain as determined by Raman measurements. We suggest that the proposed method is of general importance for obtaining the NRG on any support, which is of significance for applications, *e.g.* devices where uniform, uniaxial rippling of graphene is desirable.

**2. Experimental**

2.1 Sample preparation.

Ir(332) was cleaned by argon ions sputtering, oxygen burning and subsequent annealing and slow cooling [37]. Single layer graphene was grown by using procedures described in reference Šrut *et al.* [37]. Specifically, samples were prepared via chemical vapor deposition in an ultra high vacuum (UHV) with an Ir crystal at temperature of 1080 K exposed to an ethylene pressure of $1\times10^{-7}$ mbar. For intercalating graphene, Cs was deposited by using commercially available alkali metal dispensers (SAES Getters) operated at the typical flux of $\sim 10^{15}\ m^{-2}s^{-1}$. During Cs deposition, the sample was kept at room temperature.

2.2 Graphene transfer procedure

The schematic representation of the graphene lift-off and transfer process is depicted in Figure 1a - d with the photograph of each transfer step shown on Figure 1e - h. Prior to removal from UHV, graphene on Ir(332) was first intercalated with Cs. The crystal was then taken out of the UHV chamber (cf. Figure 1e) and its surface was coated with a drop of 2% solution of polycarbonate in anisole (cf. Figure 1a). The sample was then dried until the polycarbonate solidified forming a reinforcing thin film on the graphene. Such sample was subjected to a so-called bubbling transfer. This method was previously successfully applied for graphene transfer from Pt and Ru substrates [35,36]. In our case, the coated crystal was immersed in a 1N NaOH solution and was acting as the cathode while a piece of Pt foil served as an anode (Figure 1b and f). The electrochemical process was divided into two steps. First, a voltage of around 1.2 V was applied which is sufficient to induce so-called under potential deposition/intercalation of hydrogen between graphene and Ir [36]. We believe that this process is also accompanied by intercalation of water molecules [38] along with hydrogen (Figure 1c). The macroscopic edge of the intercalation front is visible under optical microscope as interference lines in Figure 1g, marked with an arrow in the magnified inset. Second step of the process includes application of higher voltage (around 2-3 V) above the threshold for hydrogen evolution where large bubbles of $H_2$ gas lift-off the graphene together with polycarbonate layer away from the Ir substrate. Graphene with polycarbonate was subsequently washed in deionized water and transferred onto a Si wafer terminated by 300 nm thick layer of $SiO_2$. Polycarbonate was then removed by washing the sample with dichloromethane. The final result of the transfer process, shown in Figure 1d and h, is a graphene sheet of 6 mm in diameter on Si/SiO$_2$ visible with a naked eye [39].

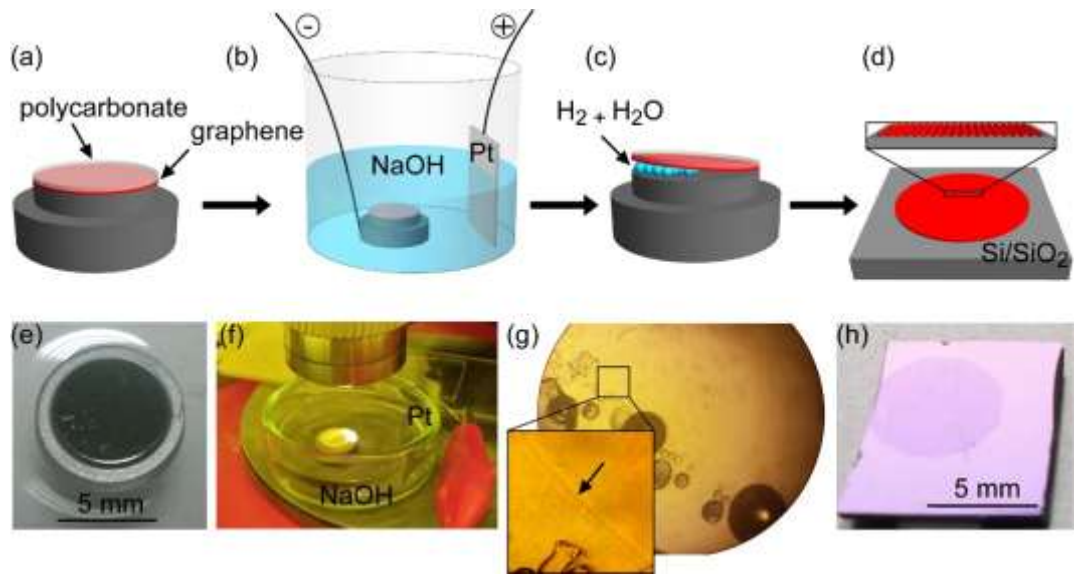

Figure 1. (a) – (d) Schematic representation of the graphene transfer procedure steps described in the text. (e) Photograph of the Ir(332) crystal covered with graphene monolayer after the sample has been taken out of UHV. (f) Photograph of an experimental setup for the bubbling transfer. (g) Optical microscopy image (80x magnification) of the sample during the under-potential treatment. Inset shows the magnified region marked by a black rectangle where the intercalation front indicated by a black arrow can be seen. (h) Photograph of a graphene sheet after the transfer to Si/SiO$_2$.

The above described bubbling transfer can also be applied to graphene on Ir(332) without any prior intercalation in UHV, however, we find it usually results in macroscopically visible incomplete transfer, where either severed graphene flakes or just the outer rim of the sample is transferred (not shown). During the optimization of the transfer procedure, we found that complete graphene sheets with less lift-off damage can be achieved when the sample is intercalated prior to the transfer, which can be effectively done by Cs (cf. Figure 2f). It appears that this step opens door for efficient under potential intercalation of hydrogen and water in the electrochemical step. The transfer process takes less than an hour from the point of removing the sample from UHV.

2.3 Experimental techniques

The low energy electron diffraction (LEED) patterns were measured from samples kept at room temperature using 4-Grid SpectaLEED (Omicron). The scanning tunneling microscopy (STM) measurements were performed by using Aarhus STM (SPECS). The STM data was measured with sample kept at room temperature, with the tip grounded and the sample put to a bias voltage. All of the atomic force microscopy (AFM) measurements were performed with Flex AFM (Nanosurf) in contact mode with the set point force 10 nN. Used AFM tips were Nanosensors (PPP-CONTR-10) with a force constant 0.02 - 0.77 N/m and a tip radius of curvature < 10 nm. All of the STM and AFM data analysis was performed in the WSxM software [40].

The micro-Raman polarized spectroscopy measurements of the graphene samples transferred to a wafer surface were performed using a Horiba Jobin-Yvon T64000 system, with a resolution of 0.1 cm$^{-1}$ and equipped with an argon ion laser for the excitation, operating at 514.5 nm (Coherent Innova 400). The polarization and power of the incident light were tuned by the half-wave plate and filters. The laser power of 1 mW was focused on the graphene sample in order to prevent known heating effect of the sample, using a ×50 objective lens. The polarized Raman spectroscopy measurement were performed by the half wave plate for 180°, in steps of 30°. To obtain a Raman spectra with good statistics the acquisition time was set to 60 s per spectrum with 10 repeats.

## 3. Results and discussion

Synthesis of a single layer graphene on Ir(332) leads to inevitable substrate restructuring during the growth [37]. The surface restructuring results in an alternating array of terraces and step bunches with step edges following the original direction of the step edges on the clean substrate in the [$\bar{1}10$] direction [41]. This property is not sensitive to the orientation of graphene overlayer with respect to the Ir substrate, which on the other hand can be varied depending of the applied growth conditions [37,41]. Specifically, several different dominant graphene orientations on Ir(332) can be achieved [37,41] ranging from graphene aligned with the substrate (0° rotated, noted R0), up to the one rotated by 30° (noted R30) [42,43]. Figure 2a shows the LEED pattern from graphene on Ir(332) where one Ir spot is marked with the red arrow, and the R0 and R30 graphene spots are marked with the green and red circle, respectively. In addition, diffraction spots of the R26, mirrored R34 and R8 graphene are also visible (yellow, blue and magenta circle, respectively). The dominating graphene contributions are R30 and R26/R34, which is visible from the pronounced intensity of the corresponding LEED spots (cf. Figure 2a). The applied growth procedure, yielding a high ratio of the R30 and close to R30 graphene, was chosen because such graphene samples exhibit pronouncedly strait step edges, which are sharper and thus better defined than in the case of graphene with rotations grouped around R0 [37]. Figure 2b shows a differentiated scanning tunneling microscopy (STM) image of the same sample. Dark stripes are associated with (111) terraces and bright stripes with (331) step bunches (unpublished work). The Fourier transform (FT) of the STM image in Figure 2b is displayed in Figure 2c. The FT image is characterized by a linear intensity distribution indicating a presence of the aligned 1D features parallel to the [$\bar{1}$10] direction, which is also obvious from the real space STM image. Additionally, the FT image shows two pronounced spots along the linear intensity distribution, which correspond to the periodicity of 13.3 nm. We attribute this value to the characteristic structural motif of repeating terraces and step bunches. Figure 2d upper panel shows STM topograph of a terrace and step bunch motif with denoted facet orientation. In this case, the motif is 40 nm wide and reaches the amplitude of 3 nm, as can be seen from the line scan on the Figure 2e. The terrace and step bunch display a characteristic sawtooth profile. Finally, before the sample was removed from the UHV, it was intercalated with Cs. The intercalated amount corresponds to a saturated monolayer of Cs packed in a $(\sqrt{3} \times \sqrt{3})R30°$ structure relative to Ir (111) [44]. This can be clearly seen in the LEED pattern in Figure 2f, where one Cs diffraction spot was marked by a light blue circle and two Ir diffraction spots were marked by red arrows. According to STM characterization, the Cs intercalation did not affect the graphene morphology on the Ir(332), similarly to the case of Cs intercalated graphene on Ir(111) [44].

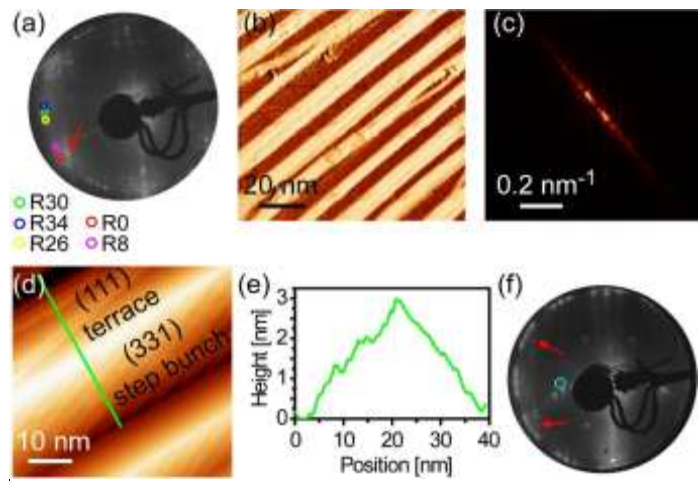

Figure 2. (a) LEED pattern of gr/Ir (332) taken at an electron energy of 75 eV. (b) Differentiated STM image showing sample after graphene synthesis. (c) Fourier transform image of (b). (d) STM topograph revealing characteristic terrace and step bunch motif. (e) The line profile taken along the green line in (d). (f) LEED pattern of gr/Ir (332) intercalated with 1 ML of Cs taken at an electron energy of 56 eV.

Structure of the gr/Ir(332) sample in real space over larger scales can be obtained from the atomic force microscopy imaging. The lateral resolution of AFM is in our case limited by the curvature of tips used (5 - 10 nm). The AFM characterization of gr/Ir(332) was performed after the samples were extracted from UHV to atmosphere by taking images on several different locations of the sample separated typically by several mm (c.f. Figure 3a - c). All AFM images were recorded by using the same scanning direction and they reveal that all imaged 1D structures follow the same direction. This implies that the rippling of our sample is macroscopically uniform in the $[\bar{1}10]$ direction of Ir step edges. The surface roughness analysis of Figure 3a - c showed that these features have an average roughness of 3.1 nm. The FT image of the AFM topograph from Figure 3b is presented in Figure 3d. Same as in the case of periodic features from STM image FT in Figure 2c, Figure 3d consists of a single intensity line confirming the 1D-ordering at the surface. Moreover, the FT now shows two sets of relatively faint spots corresponding to periodic features of characteristic lengths of about 38 nm and 57 nm. In the real space image this is resolved as a periodic repetition of higher (brighter) areas with larger separation and lower (darker) more closely spaced areas, respectively.

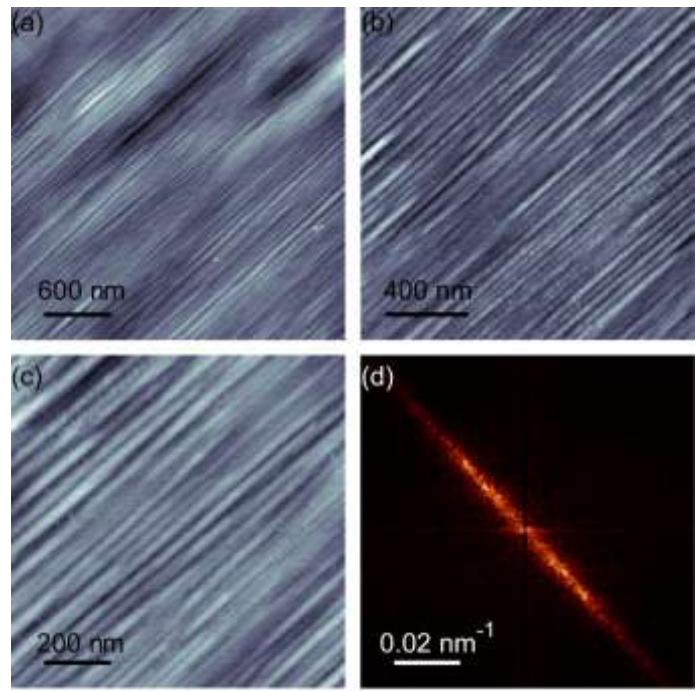

Figure 3. (a) - (c) AFM topograph of the same sample recorded at three different positions using the same scanning angle. (d) Fourier transform of the image shown in (b).

Graphene transferred to Si/SiO$_2$ was subjected to AFM imaging and Raman spectroscopy characterization. The AFM topographs acquired at several different locations on the sample separated typically by several mm are presented in Figure 4a - d. It is evident that the graphene exhibits a 1D ripple structure with a well-defined uniform rippling direction. The FT in Figure 4e shows again strait intensity line, now with two distinct spots. The strait line is a result of a 1D ripples ordering while the observed spots indicate an average periodicity of the ripples of 67 nm. The surface roughness analysis of Figure 4a - d showed that the observed roughness ranges from 0.4 nm to 1.7 nm, and it averages around 1.1 nm. The AFM line profile along the green line on Figure 4d is presented in Figure 4f. The periodicity of the ripples analyzed in this particular line scan was 40 nm and the average roughness was 0.47 nm.

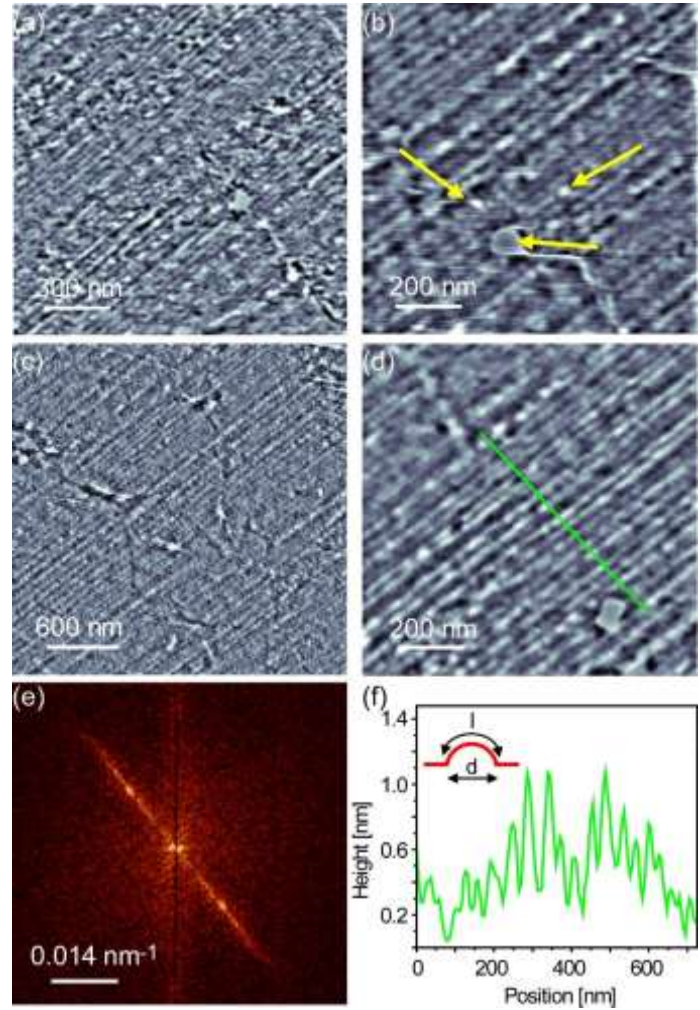

Figure 4. (a) – (d) AFM topographs at several different locations at the same sample (cf. Figure 1h). (e) Fourier transform image of (c). (f) AFM line profile corresponding to the green line in (d). Inset shows a simplified ripple cross-section used for strain calculation.

Under simplified assumption that the characteristic ripple cross-section follows a circular segment (cf. inset on Figure 4f), the strain $\varepsilon$ in graphene sample can be calculated trough a simple formula $\varepsilon = (l - d)/d$ where l is a length of a circular segment and the d is a width of a ripple. From Fig. 4f, the average strain in the ripples is calculated to be 0.037 %. Calculations carried out by taking the periodicity acquired the FT (Figure 4f) and height from roughness analysis give the higher value of 0.072 %. We believe the higher value is overestimated especially due to impurities affecting the average height.

Furthermore, transferred samples exhibit a small number of defects, cracks and tears, clearly visible in Figure 4a – d. It is likely that these defects were created during the lift-off or transfer as they were not observed in AFM imaging of graphene on Ir(332) before the transfer (cf. Figure 3a - c). In addition to cracks, the on top contaminations are observed (cf. Figure 4f indicated by yellow arrows), probably polycarbonate residues which remained after the sample cleaning. Such features affect the AFM scanning stability by increasing the level of noise in the images.

Raman spectroscopy is a particularly good method to study both graphene and other carbon based materials properties [45]. The Raman spectrum recorded using an unpolarized laser beam is presented in Figure 5a. Two distinct graphene peaks, the 2D at 2694 cm$^{-1}$ and the G at 1587 cm$^{-1}$, are clearly visible and also a small contribution from the D peak at 1347 cm$^{-1}$ [46,47]. Each peak was fitted by a single Lorentzian curve and the corresponding full width at half maximum (FWHM) of the 2D, G and D peaks was 34 cm$^{-1}$, 17 cm$^{-1}$ and 13 cm$^{-1}$, respectively. The peak positions and the extracted FWHM's indicate that the transferred graphene flake was indeed a monolayer graphene [47–49]. Please note that the FWHM of the 2D peak that we measure is on the higher end of the values range reported in the literature for the monolayer graphene [49]. This is likely a consequence of the appearance of the D peak *i.e.* a result of present defects and/or edges [46,50] in the graphene sample, which is not surprising due to cracks and tears observed in AFM topographs (Figure 4a – d) and the fact that graphene grows in several different orientations (the inevitable presence of grain boundaries). The D and G peak intensity ratio is, however, relatively small $I_D/I_G = 0.127$. Taking into account a simplified assumption that the D peak contribution comes from point defects [51], one can easily calculate the defects density of $3.26 \times 10^{10}\ cm^{-2}$, or 1 point defect per every $1.2 \times 10^5$ C atoms.

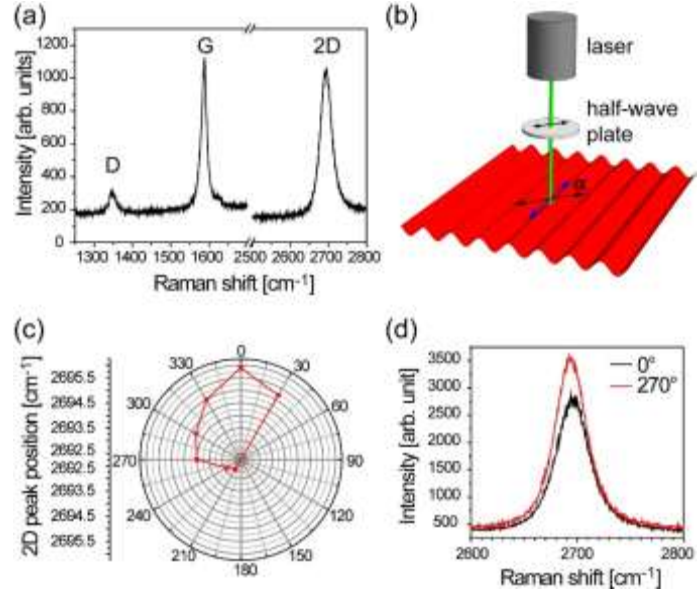

Figure 5. (a) Raman spectrum of the graphene sample on $SiO_2/Si$ recorded with an unpolarized laser light. (b) A schematic model of the polarized laser light Raman measurement. Black arrows mark laser polarization and red arrow marks a direction of the graphene ripples. (c) Polar plot showing positions of the 2D peak with polarized light for different laser polarization angle. (d) Raman spectra of two 2D graphene peaks for two different angles of light polarization separated by 90°.

Furthermore, the Raman peak positions of the G and 2D peaks in graphene are known to depend on a number of factors including strain, in which case a redshift of both peaks is expected to take place [20]. For the uniaxially strained graphene it is possible to separately detect Raman peaks corresponding to strained and unstrained directions, or any direction in between. This can be done by using a polarized laser beam where the position of both the G and 2D peaks depend on the angle of the polarized light with respect to the direction of the applied strain [20,21]. Schematic model for such Raman setup which we used to characterize the strain in transferred sample is shown on Figure 5b, where $\alpha$ denotes an angle between laser polarization (black arrow) and ripple direction (blue arrow). Aligning the laser polarization with the direction of the applied strain (perpendicular to the ripples *i.e.* $\alpha = 90°$), only strained graphene direction is probed and the position of, for example, 2D peak should be maximally redshifted at its minimum wave number. Conversely, by rotating the light polarization by 90° ($\alpha = 0°$), the unstrained graphene direction is probed and the 2D peak position should be at the maximum wave number. The measurement is always preformed on the same spot on the sample, as we rotate the light polarization using a half-wave plate. The polarized Raman measurements performed here at the same time eliminate other contributions which may affect the 2D peak shift, primarily the doping. For expected isotropic doping in the area characterized by Raman, the change in the light polarization does not have any effect. We have performed a quantitative analysis of strain via Raman measurements with polarized laser light and the results are shown in Figure 5c. The position of the 2D peak is presented in a polar plot, where measurements were done in 30° steps. A characteristic leaf-like shape of the 2D peak positions [21] is visible, confirming the presence of a well-defined uniaxial strain in the sample. Figure 5d shows an overlay of two different 2D Raman peaks measured with a difference in polarization angle of 90°. The position difference of these two peaks is 1.92 cm$^{-1}$. Comparing this value to a systematic characterization of Mohiuddin *et al.* [20], a strain of 0.03 % is determined, which is in excellent agreement with the strain obtained from the simple model based on AFM measurements described above.

We have shown that prior to the transfer, gr/Ir(332) has a well-defined step edge orientation in the [$\bar{1}$10] direction, visible both in STM and AFM measurements (Figure 2b and Figure 3a - c). However, the FT of STM and AFM images revealed different periodic contributions, namely the FT of a STM image showed periodicity of 13.3 nm, and the FT of an AFM image showed two periodicities of 38 nm and 57 nm. One of the reasons for this difference likely comes from the fact that the AFM scanned larger areas than the STM, in which case local feature size variations of terraces and step bunches may lead to differences. In our previous paper we have shown that gr/Ir(332) displays certain terrace width distribution [37]. For example, the individual terrace - step bunch motif resolved by STM in Figure 2d is wider than the average terrace - step bunch width obtained by analyzing the Figure 2b. Additionally, this individual terrace - step bunch motif on Figure 2d is of the same size as the structures resolved in Figure 3a - c. Moreover, probably more important reason for the observed different periodicities in STM and AFM is the fact that AFM tip has a finite radius of about 5 - 10 nm, which limits AFM lateral resolution and smoothens sharp edges that STM easily resolves. This is readily seen in Figure 3c - d where the observed periodic features appear round, however from the STM images we know that the periodic terrace – step bunch motif has in fact a sawtooth profile (cf. Figure 2d). Thus, we propose that recorded AFM images likely show a superposition of several terraces and step bunches, as well as some individual larger features (such as the one from Figure 2d).

Our characterization indicated that the fast transfer process resulted in a millimeter scale macroscopic monolayer graphene, where the size of the graphene flake is only limited by the size of the Ir crystal (cf. Figure 1e and h). The repeated transfer process does not affect the quality of the substrate Ir crystal which can be used again and again to synthesize and transfer graphene. The important step to perform a successful transfer was facilitated through Cs intercalation under UHV, which drastically reduced the lift-off damage. This is due to the fact that the intercalation of Cs atoms weakens graphene - Ir binding interaction, primarily by increasing the distance between the graphene and Ir [44], which apparently helps to facilitate electrochemical process. Similar reasoning was applied before by Herbig *et al.* who used $Br_2$ intercalation to mechanically exfoliate graphene from flat Ir substrate Ir(111) [52], and more recently Koefoed *et al.* used intercalation of large organic tetraoctylammonium ions to reduce graphene - Ir(111) binding and subsequently transfer the graphene [53].

Graphene transferred from Ir(332) onto Si/$SiO_2$ showed macroscopic, well ordered, periodic, uniaxially rippled structure of nanometer dimensions which was readily seen in AFM measurements (cf. Figure 4). Periodicity of ripples ranged between 40 nm and 70 nm, and their corresponding roughness between 0.4 nm and 1.7 nm. The average ripple size was 1.1 nm and the periodicity 67 nm. The ripples extended in the same direction over the entire sample matching the [$\bar{1}$10] direction of the step edges on the Ir(332) surface. Due to the ripples uniaxial arrangement and their characteristic size they can not be a consequence of the Si/$SiO_2$ substrate surface small microscopic corrugation which is not periodic and has a characteristic length scale of about 10 nm in all directions [32]. The AFM characterization indicates that the transferred NRG has approximately the same periodicity as gr/Ir(332) before the transfer but it displays on average 3 times lower corrugation. We speculate that the smaller corrugation comes from the relaxation of graphene on a flat wafer. We suggest that the characteristic profile changes from a sawtooth-like shape to a more rounded profile to reduce bending strain, with bottom parts of ripples laying flat at the surface due to adhesion with the flat wafer substrate [54]. Nevertheless, the macroscopic uniformity of the rippled structure resulted in the uniaxial strain in the transferred NRG as confirmed by the polarized Raman spectroscopy measurements (cf. Figure 5c and d). The strain values extracted from both the AFM and polarized Raman measurements match thus giving a credible estimate of the inherent uniaxial strain present in the NRG system, which averages to be around 0.034 %.

Crucial on the way to obtain the NRG on any support, *e.g.* dielectric, is the quality of lift-off from iridium and transfer. In our case, we have coated graphene for the transfer process with a polycarbonate, as it provides a stiffer support for graphene and results in less impurities in graphene compared to more commonly used PMMA. Nevertheless, our results indicate that the transfer leaves some defects and cracks in the graphene sheet. They are directly imaged by AFM (Figure 4a – d) and fingerprinted in Raman spectroscopy trough the appearance of the D peak (Figure 5a). In addition partial contribution to the D peak could also come from the grain boundaries which are intrinsically present due to graphene rotational domains [50,55,56]. Raman measurements of the graphene on Ir(332) could distinct between the two contributions of the D peak, however, similarly as for graphene on Ir(111), the graphene-Ir hybridization fully suppresses the Raman active modes [57], which prevented us from recording Raman spectra prior to transfer from Ir. Finally, leftover contamination is present on the top of graphene from residual polycarbonate (cf. Figure 4b), which may affect the sample quality estimates based on Raman spectra. Please note that the measured D and G peak intensity ratio in our case is two times smaller then what was reported for graphene transferred from Ir(111) by similar method [53]. Specifically, the number of point defects that we calculated above turns out to be two orders of magnitude smaller than in the work of Koefoed *et al.* [53]. We believe that such large discrepancy comes from the fact that two different ways to calculate the number of point defects have been applied. In any case, more reliable way to compare samples transferred from flat Ir(111) [53] and in our case from Ir(332) should refer to the $I_D/I_G$ ratio. Nevertheless, a smaller D to G peak intensity ratio for graphene transferred from Ir(332) is surprising because intrinsic presence of defects (more rotational grain boundaries) should favor the quality of graphene from Ir(111). This indicates that apparently the quality of the transfer process presented in our work is less invasive regarding the final quality of transferred graphene.

The technique which was applied here for the system graphene on the stepped Ir for the first time, ensures large scale, fast transfer of graphene preserving at the same time periodic nano-ripple morphology. The preservation of the rippled morphology was expected, especially in the light of the work of Kraus *et al.* [58] who showed that, even after etching away a part of the Cu substrate, the resulting freestanding graphene keeps its faceted structure over micrometer scale. Additionally, several different works showed that graphene transfer from faceted, or in other way structured surface yields a NRG [27,28,33]. However, all of the previously reported NGRs possess widely spaced ripples mostly in μm range, which are non-periodic or quasi-periodic, and usually have no orientation control. The advantage of using Ir(332) as a substrate for growing the NRG lies in the fact that growth parameters can be used to selectively tune specific graphene rotations (cf. Figure 2a) which all bend across the same substrate step edge which runs in the [$\bar{1}$10] direction [37,41]. This then allows a production of NRG with tailored strain directions (zig-zag, armchair, or any arbitrary direction in between), which is significant for possible applications in straintronics. By changing the temperature it is also possible, to a certain extent, to adjust the widths of the (111) terraces and (331) step bunches [37], making the growth temperature a tool to affect the desired width of the transferred graphene ripples. Finally, based on the AFM and Raman analysis our experiments leave an ample space for the improvement of the transferred NRG through the optimization of transfer parameters and more detailed cleaning procedures. Moreover, they open possibility for additional characterization, namely systematic transport measurements.

## 4. Conclusions

We have successfully performed fast large scale transfer of a millimeter sized single layer graphene from Ir(332) to a dielectric support. A single crystal Ir can be repeatedly used for graphene synthesis and transfer, and the size of the crystal is the only limiting factor for the size of the transferred graphene. Graphene growth on Ir(332) results in a surface faceting creating alternating (111) and (331) facets where the step direction is preserved over the entire sample, resulting in the periodic nano-rippled graphene overlayer. After the transfer to a flat dielectric $Si/SiO_2$ support, graphene preserves a periodic nano-ripple morphology with macroscopically well-defined ripple direction. Such NRG exhibits the uniaxial strain of 0.03% which is measurable on a macroscopic scale using polarized Raman spectroscopy and more roughly by the AFM topography characterization. Due to possibilities of tuning graphene synthesis on Ir(332) via growth parameters, this NRG system offers an opportunity to select a desired strain direction with respect to graphene lattice directions, as well as tuning the geometry and periodicity of the ripples. Both of these features are potentially useful for straintronic applications.

**Acknowledgments**

I.Š.R. acknowledges funding from the foundation L'Oréal Adria - UNESCO, For women in science. The support through the Center of Excellence for Advanced Materials and Sensing Devices, research unit for Graphene and Related 2D Structures is gratefully acknowledged.